\documentstyle[12pt]{article}
\def\fzweinu{F_2^{\stackrel{(-)}{\nu}N}}
\def\fzweic{{\cal{F}}_2^{\stackrel{(-)}{c}}}
\begin{document}
\setlength{\baselineskip}{0.75cm}
\setlength{\parskip}{0.45cm}
%%%%%%%%%%%%%%%%%%%%%%%%%%%%%%%%%
% TITLEPAGE
%%%%%%%%%%%%%%%%%%%%%%%%%%%%%%%%%
\begin{titlepage}
\begin{flushright}
DO-TH 96/06 \linebreak
March 1996
\end{flushright}
\vskip 1.2in
%\vskip 0.5in
\begin{center}
{\Large\bf The Strange Sea Density and Charm Production in Deep
Inelastic Charged Current Processes}
\vskip 0.8in
{\large M.\ Gl\"{u}ck, S.\ Kretzer and E.\ Reya}
\end{center}
\vskip .3in
\begin{center}
{\large Institut f\"{u}r Physik, Universit\"{a}t Dortmund \\
D-44221 Dortmund, Germany}
\end{center}
\vskip 1in
{\large{\underline{Abstract}}}

\noindent
Charm production as related to the determination of the strange sea
density in deep inelastic charged current processes is studied 
predominantly in the
framework of the $\overline{{\rm{MS}}}$ fixed flavor factorization 
scheme.
Perturbative stability within this formalism is demonstrated. The
compatibility of recent next-to-leading order strange quark
distributions with the available dimuon and $F^{\nu N}_2$ data is
investigated. It is shown that final conclusions concerning these
distributions afford further analyses of presently available and/or
forthcoming neutrino data.
\end{titlepage}
%
%%%%%%%%%%%%%%%%%%%%%%%%%%%%%%%%%%%%
%
\noindent
Heavy quark production at high energy neutral current reactions was
recently shown \cite{ref1} to be optimally described within the
framework of a fixed flavor (factorization) scheme (FFS) where, besides
the gluon $g$, only the $u$, $d$ and $s$ quarks are considered as
partons and any heavy quark ($c$, $b$, ...) contribution is calculated
in fixed order $\alpha_s$ perturbation theory. Within this framework the
charged current production of a heavy quark pair such as $t \bar{b}$ in
$\nu p\rightarrow \mu^-\ t \bar{b}\ X$ follows the same pattern \cite{ref2}
utilizing the relevant formulae for the underlying '$W g$ fusion'
subprocess $W^+ g \rightarrow t \bar{b}$. Since both $m_{t,b} \gg
\Lambda_{QCD}$, we do not encounter any mass singularities here and a
treatment within the framework of the FFS is straightforward and
unproblematic. This favorable situation changes, however, when
considering the corresponding charm production process (e.g.\ $W^+ g
\rightarrow c \bar{s}$) since the associated strange quark is taken as
massless in the FFS and considered as a parton. In contrast to the
former cases we encounter here a mixed situation which affords a careful
treatment within the framework of the FFS. The leading order (LO)
contribution for charm production in $\nu N \rightarrow \mu^-\ c\ X$,
$N = (p+n)/2$, comes from the basic ${\cal{O}}(\alpha_s^0)$
subprocess $W^+ s' \rightarrow c$ where
\begin{equation}
s_{\nu N}'\ \equiv\  \left|V_{cs}\right|^2 s\ +\ \left|V_{cd}\right|^2
\frac{d+u}{2}
\end{equation}
with $\left|V_{cs}\right| = 0.9743$ and $\left|V_{cd}\right| = 0.221$.
The $W^+ g \rightarrow c \bar{s}$ fusion process yields the essential
part of the next-to-leading order (NLO) correction where the other part
is due to the subprocess $W^+ s' \rightarrow g c$. Both subprocesses
posses a mass singularity associated with \mbox{$m_s = 0$} which is absorbed
via dimensional regularization into the renormalized, 
\mbox{$Q^2$--dependent},
parton distribution $s'$. The remaining finite pieces of $W^+ g
\rightarrow c \bar{s}$ and \mbox{$W^+ s' \rightarrow g c$} then yield the
genuine NLO correction to $\: W^+ s' \rightarrow c\:$ which will henceforth
be considered in the now commonly adopted $\overline{\rm{{MS}}}$ factorization
scheme. Denoting the contributions of the above subprocesses to the
structure functions $F_i(x,Q^2)$ by $F_i^c(x,Q^2)$ and defining
furthermore ${\cal{F}}_1^c \equiv F_1^c$, ${\cal{F}}_3^c \equiv
F_3^c/2$, ${\cal{F}}_2^c \equiv F_2^c/2\xi$ where $\xi=x(1+m_c^2/Q^2)$,
one obtains in NLO \cite{ref3}
\begin{eqnarray}
{\cal{F}}_i^c(x,Q^2)\ \  =\ \  s'(\xi,\mu^2) &+&
\frac{\alpha_s(\mu^2)}{2\pi}\left\{\int_{\xi}^1 \frac{d\xi '}{\xi '}
\left[H_i^q(\xi ',\mu^2,\lambda)\ s'(\frac{\xi}{\xi '},\mu^2)
\right. \right. \nonumber\\
&+& \left. \left. H_i^g(\xi ',\mu^2,\lambda)\ g(\frac{\xi}{\xi '},\mu^2)
\right]\right\}\ \ .
\end{eqnarray}
Here $\lambda \equiv Q^2/(Q^2+m_c^2)$ and the $H_i^{q,g}$ are given, up
to minor modifications specified in the Appendix, in ref.\ \cite{ref3}.
The specific choice for the factorization scale $\mu$ will be studied
below. The inclusive cross section in terms of the $F_i(x,Q^2)$ is given
by
\begin{equation}
\frac{d^2\sigma^{\nu(\bar{\nu})}}{dx\ dy}\ =\ \frac{G_F^2 M_N
E_{\nu}}{\pi(1+Q^2/M_W^2)^2}\
\left[(1-y)F_2^{\nu(\bar{\nu})}+y^2xF_1^{\nu(\bar{\nu})}\pm
y(1-\frac{y}{2})xF_3^{\nu(\bar{\nu})}\right]\ \ .
\end{equation}

To study the size of the NLO corrections we shall utilize the LO and NLO
parton distributions of \cite{ref4} which are already conceived in the
FFS being furthermore $\overline{\rm{{MS}}}$ distributions in the NLO. In
addition we shall also employ the LO and NLO($\overline{\rm{{MS}}}$) 
\mbox{CTEQ3 \cite{ref5}} 
and the NLO($\overline{\rm{{MS}}}$) MRS(A) \cite{ref6} parton
densities which refer to the 'variable flavor' scheme where intrinsic
charm densities are purely radiatively generated using the ordinary
massless evolution equations, starting at $Q = m_c$. For definiteness we
show in fig.\ 1 the quantity
\begin{equation}
\xi s(\xi,Q^2)_{eff}\ \equiv\ \frac{1}{2}\
\frac{\pi(1+Q^2/M_W^2)^2}{G_F^2M_N E_{\nu}}\ \left|V_{cs}\right|^{-2}\
\frac{d^2\sigma^{(c \bar{s})}}{dx\ dy}
\end{equation}
which has been also studied experimentally \cite{ref7} and where the
superscript $c \bar{s}$ refers just to the CKM non-suppressed ($V_{cs}$)
component of $s'$ in eqs.\ (1--3). Note that in LO the cross section in (4)
reduces to
\setcounter{equation}{3}
\def\theequation{\arabic{equation}'}
\begin{equation}
\xi s(\xi,Q^2)_{eff}\ =\ (1-\frac{m_c^2}{2 M_N E_{\nu} \xi})\ \xi
s(\xi,\mu^2) + {\cal{O}}(\alpha_s)\ \ \ .
\end{equation}
\def\theequation{\arabic{equation}}
As can be seen in fig.\ 1 the NLO corrections to the LO results are
reasonably small and, in particular, do \underline{not} afford a drastic
change of $s(x,Q^2)$ when passing from the LO to the NLO analysis.

This contrasts with the conclusions of the CCFR group \cite{ref7} whose
NLO $s(x,Q^2)$ is almost twice as large as compared to their previous
\cite{ref8} LO $s(x,Q^2)$. The analysis of the CCFR group is based on
the NLO formalism of \cite{ref9} which is not strictly equivalent to our
NLO($\overline{{\rm{MS}}}$) FFS formalism but still is expected to yield quite
similar results. The enhancement of the NLO $s(x,Q^2)$ in \cite{ref7}
can therefore not be attributed to the different formalism itself
\cite{ref10} but rather to its inconsistent application. In the
formalism of \mbox{ref.\ \cite{ref9}} one considers the $W^+ g \rightarrow c
\bar{s}$ contribution with $m_s \neq 0$, i.\ e.\ employs a finite mass
regularization and subtracts from it that (collinear) part which is
already contained in the renormalized, $Q^2$--dependent $s(x,Q^2)$. The
CCFR group applied their acceptance corrections to the full contribution
from $W^+ g \rightarrow c \bar{s}$, which corresponds effectively to a
multiplication with an acceptance correction factor ${\cal{A}} = 0.6 \pm
0.1$, while inconsistently keeping the subtraction term in its full
original (acceptance \underline{un}corrected) strength \cite{ref11}.
This latter subtraction term is given, relative to $\xi s(\xi,\mu^2)$, 
by
\begin{equation}
{\rm{SUB}}\ \equiv\ \frac{\alpha_s(\mu^2)}{2\pi}\ \ln\frac{\mu^2}{m_s^2}\
\int_{\xi}^1 \frac{dz}{z}\ g(z,\mu^2)
\ P_{qg}^{(0)}\left(\frac{\xi}{z}\right)
\end{equation}
using \cite{ref11} $m_s=0.2\ {\rm{GeV}}$. In fig.\ 2 we compare the
result obtained in this manner with the one where also the subtraction
term \cite{ref9} in (5) was consistently multiplied by the same
acceptance factor ${\cal{A}}$. The result corresponding to the acceptance
uncorrected subtraction term (dashed curve) clearly demonstrates that
SUB alone [eq.\ (5)] represents too strong a suppression of the mass
singularity component in ${\rm{LO}}+{\cal{A}}\ast {\rm{NLO}}$ and that the
correct result (solid curve) in fig.\ 2 is almost a factor of 2 larger
in the small-$x$ region. This implies that instead of $s_{{\rm{NLO}}}
\simeq 2\ s_{{\rm{LO}}}$ for
$x \sim$ \hspace{-0.5cm}\raisebox{1ex}{$<$}
$0.1$, as inferred by CCFR
\cite{ref7} and used for our analysis in fig.\ 2, one rather needs a
smaller $s_{{\rm{NLO}}}$, i.\ e.\ closer in size to $s_{{\rm{LO}}}$, in
order to \underline{reduce} the solid curve in fig.\ 2 and to bring it
closer to experiment. Here we have chosen~\cite{ref7} a factorization
scale $\mu=2\ p_T^{{\rm{max}}}=\Delta (W^2,m_c^2,M_N^2)/W$, where
$p_T^{{\rm{max}}}$ is the maximum available transverse momentum of the
final state charm quark; the results in fig.\ 2 remain practically
unaltered with the alternative choice $\mu^2=Q^2+m_c^2$.

A further feature emanating from the fits in \cite{ref7} was
$m_c^{{\rm{NLO}}} \simeq 1.7\ {\rm{GeV}}$ as compared to the previous
\cite{ref8} $m_c^{{\rm{LO}}} \simeq 1.3\ {\rm{GeV}}$ which further
suppressed the NLO cross section and demanded the unusual, even more
enhanced NLO $s(x,Q^2)$. A consistent treatment of the acceptance
correction would most probably also lower the fitted $m_c^{{\rm{NLO}}}$
down to a more reasonable $m_c^{{\rm{NLO}}} \simeq 1.5\ {\rm{GeV}}$ and
bring the NLO $s(x,Q^2)$ close to the LO $s(x,Q^2)$.

Our conclusions concerning the strange quark distributions of
\cite{ref4,ref5} are that they agree in LO with \cite{ref8} and are not
refuted in NLO by the analysis in \cite{ref7}. Furthermore due to the
perturbative stability demonstrated in fig.\ 1 we expect the NLO strange
quark distributions of [4--6] to lie in the correct ball park.
For a final conclusion concerning these matters, a reanalysis of
presently available dimuon neutrino data is obviously mandatory!

It is also interesting to check the above statements by an independent
quantitative test sensitive to $s(x,Q^2)$ such as for example the
combination $\frac{5}{6} F_2^{\nu N} - 3 F_2^{\mu N}$ which is given
approximately by
\begin{equation}
\frac{5}{6}\ F_2^{\nu N}(x,Q^2)\ -\ 3\ F_2^{\mu N}(x,Q^2)\ \simeq
\ xs(x,Q^2)
\end{equation}
where the charm contributions, the $m_c^2/Q^2$ corrections and the NLO
$q$-- and $g$-- induced contributions are rather small and, furthermore,
$\left|V_{cs}\right|^2\simeq 1$ and $\left|V_{cd}\right|^2\simeq 0$. In
fig.\ 3a we compare various LO and NLO results for $xs(x,Q^2)$ in eq.\
(6) with the published~\cite{ref12} and more recent but 
preliminary~\cite{ref13} ${\stackrel{(-)}{\nu}} N$
data and the NMC (deuteron) $\mu N$ data~\cite{ref14}. It should be
kept in mind that the neutrino data refer to a Fe--target and are
therefore very sensitive to nuclear (EMC) corrections in the small-$x$
region: Only the preliminary (unpublished) neutrino data \cite{ref13},
which are larger than the published ones~\cite{ref12} in the small-$x$
region, disagree with the approximate predictions, eq.\ (6), in fig.\
3a. That this latter approximation is indeed sufficiently accurate is
demonstrated in fig.\ 3b where $xs(x,Q^2)$ is compared with the full NLO
result for $\frac{5}{6} F_2^{\nu N} - 3 F_2^{\mu N}$ which has to be
calculated in the following way (note that $F_2^{\nu N}$ always refers
to an average over $\nu$ and $\bar{\nu}$). In the FFS we have
\begin{eqnarray} \nonumber
\fzweinu (x,Q^2)\ = \
x\ \sum_{q=u,d}\ \Bigg\{(q'+\bar{q}')(x,Q^2)\ +\ \frac{\alpha_s(Q^2)}
{2\pi}\ \bigg[\Big((q'+\bar{q}')
*C_2^q\Big)(x,Q^2) \\
+\ 2\ \left(g*C_2^g\right)(x,Q^2)\bigg]\Bigg\}\ +\
2\ \xi\fzweic(x,Q^2)
\end{eqnarray}
with $q'=\frac{1}{2}(1+|V_{ud}|^2)q+\frac{1}{2}|V_{us}|^2s$, using
$|V_{ud}|^2+|V_{us}|^2=1$ with $|V_{ud}|^2=0.9743$, and where
${\cal{F}}_2^c$ is given in eq.\ (2) with the replacement $s'
\rightarrow \frac{1}{2}(s'+\bar{s}')$. The massless coefficient
functions $C_2^{q,g}$ are standard, see e.\ g.\ ref.\ \cite{ref4}, and
the convolutions are defined by
\begin{eqnarray} \nonumber
(q\ast C)(x,Q^2)\ =\ \int_x^1\frac{dz}{z}\ q(z,Q^2)\
C\left(\frac{x}{z}\right)\ \ \ .
\end{eqnarray}
The well known expression for $F_2^{\mu N}$ \cite{ref4} is
appropriately modified for an isoscalar target, with the charm
contribution $F_2^{c\bar{c}}$ calculated according to the $\gamma^{\ast}
g \rightarrow c \bar{c}$ fusion process etc.~\cite{ref1} as described,
for example, in \cite{ref4}. In the 'variable flavor' scheme
\cite{ref5,ref6}, where intrinsic charm densities $c(x,Q^2)$ are generated
radiatively by the
ordinary massless evolution equations, we have
\begin{equation}
\frac{1}{x}\ \left(\frac{5}{6}\ F_2^{\nu N}\ -\ 3\ F_2^{\mu N}\right)\ =
\ (s-c)(x,Q^2)\ +\
\frac{\alpha_s(Q^2)}{2\pi}\left[(s-c)\ast C_2^q\right](x,Q^2)\ \ \ .
\end{equation}

In view of the preliminary and contradicting nature of the
nuclear-shadowing corrected CCFR data for $\fzweinu$ used in fig.\ 3, a
decision concerning the (dis)agreement with theoretical QCD predictions
must obviously be postponed. According to our results in figs.\ 1 and 2,
implying strongly that $s_{{\rm{NLO}}}$ is similar in size to $s_{{\rm{LO}}}$,
and the ones in \mbox{fig.\ 3} which imply that the inclusion of the finite
part of $W^+ g \rightarrow c \bar{s}$ and the corresponding photon
induced $\gamma^{\ast} g \rightarrow c \bar{c}$ in conjunction with
present NLO strange quark densities [4--6] do not change significantly
the simple LO results, the theoretical predictions are rather
constrained and unique. Furthermore, the results in fig.\ 3 again
support our previous conclusions \cite{ref1} concerning the perturbative
stability \cite{ref3,ref15} of the charm production rate as calculated
in perturbative fixed order $\alpha_s$, i.\ e.\ in the FFS. A similar
analysis was carried out in \cite{ref16} where different conclusions
concerning the magnitude of the gluon induced contributions are
presented: These results are almost a factor of two larger than the full
NLO results at $x = 10^{-2}$ in fig.\ 3b since a factor of two error
seems (due to the lack of explicit formulae in \cite{ref16} it is not
possible to trace
its exact origin) to be present in the calculation of the $W^+ g
\rightarrow c \bar{s}$ contribution. Therefore, if the enhanced
preliminary $\nu N$ data at $x \sim$ \hspace{-0.5cm}\raisebox{1ex}{$<$}
$0.1$ as shown in fig.\ 3 are confirmed, the discrepancy between these
data and the rather solid and unique theoretical results, taking into
account the rather well understood dimuon data as well, will constitute
a major problem which cannot be solved within our present understanding of
the so far successful perturbative QCD.
%%%%%%%%%%%%%%%%%%%%%%%%%%%%%%%%%%
%
\newpage
\section*{Acknowledgements}
We thank A.\ O.\ Bazarko for several correspondences and private
communications. This work has been supported in part by the
'Bundesministerium f\"{u}r Bildung, Wissenschaft, Forschung und
Technologie', Bonn.
%%%%%%%%%%%%%%%%%%%%%%%%%%%%%%%%%%%%%
\newpage
\setcounter{equation}{0}
\def\theequation{A\arabic{equation}}
\section*{Appendix}

The fermionic NLO coefficient functions $H_i^q$ for heavy quark (charm)
production in \mbox{eq.\ (2)}, calculated from the subprocess $W^+ s
\rightarrow g c$, are given by
\begin{equation}
H_i^q(z,\mu^2,\lambda)\ =\ \left[P_{qq}^{(0)}(z)\
\ln\frac{Q^2+m_c^2}{\mu^2}\ +\ h_i^q(z,\lambda)\right]
\end{equation}
where $\displaystyle \quad
P_{qq}^{(0)}(z)=\frac{4}{3}\left(\frac{1+z^2}{1-z}\right)_+ \quad$ 
and
\begin{eqnarray}
h_i^q(z,\lambda)\ =\ \frac{4}{3} &\bigg\{& h^q+A_i\ \delta
(1-z)+B_{1,i}\ \frac{1}{(1-z)_+} \nonumber\\
 &+& \left. B_{2,i}\ \frac{1}{(1-\lambda
z)_+}+B_{3,i}\ \left[\frac{1-z}{(1-\lambda z)^2}\right]_+\right\}
\end{eqnarray}
with
\begin{eqnarray}  \nonumber
h^q\ =\ &-&\left(4+\frac{1}{2\lambda}+\frac{\pi^3}{3}+\frac{1+3\lambda}
{2\lambda}\ K_A\right)\delta(1-z) \\
 &-& \frac{(1+z^2)\ln z}
{1-z}+(1+z^2)\left[\frac{2\ln (1-z)-\ln (1-\lambda z)}{1-z}\right]_+
\end{eqnarray}
and
\begin{equation}
K_A\ =\ \frac{1}{\lambda}\ (1-\lambda)\ \ln (1-\lambda)\ \ \ .
\end{equation}
The coefficients in (A2) for $i=1,2,3$ are given in Table 1 where a
misprint in \cite{ref3} concerning $A_2$ was corrected.

\noindent
Table 1. Coefficients for the expansion of $h_i^q$ in (A2)\\
\begin{tabular*}{\textwidth}{@{~}l@{\extracolsep\fill}llll}
\hline\hline
$i$ & $A_i$ & $B_{1,i}$ & $B_{2,i}$ & $B_{3,i}$ \\  \hline
$1$ & $0$ & $1-4z+z^2$ & $z-z^2$ & $\frac{1}{2}$ \\
$2$ & $K_A$ & $2-2z^2-\frac{2}{z}$ & $\frac{2}{z}-1-z$ & $\frac{1}{2}$ \\
$3$ & $0$ & $-1-z^2$ & $1-z$ & $\frac{1}{2}$ \\ \hline\hline \\
\end{tabular*}

\noindent
The gluonic NLO coefficient functions $H_i^g$ for heavy quark (charm)
production in eq.\ (2), as calculated from the subprocess $W^+ g
\rightarrow c \bar{s}$, are given by
\begin{equation}
H^g_{i={1,2 \atop 3}}(z,\mu^2,\lambda)\ =\
\left[P_{qg}^{(0)}(z)\left(\pm L_{\lambda}+
\ln\frac{Q^2+m_c^2}{\mu^2}
\right)+h_i^g(z,\lambda)\right]
\end{equation}
where $\displaystyle \quad
P_{qg}^{(0)}(z)\ =\ \frac{1}{2}\ \left[ z^2+(1-z)^2 \right],
\quad L_{\lambda}\ =\ \ln\frac{1-\lambda z}{(1-\lambda)z} \quad$ and
\begin{equation}
h_i^g(z,\lambda)\ =\ C_0+C_{1,i}\ z(1-z) + C_{2,i} + (1-\lambda)\ z
\ L_{\lambda}\ (C_{3,i}+\lambda\ z\ C_{4,i})
\end{equation}
with
\begin{equation}
C_0\ =\ P_{qg}^{(0)}(z)\ \left[2\ln (1-z)-\ln (1-\lambda z) -\ln
z\right]\ \ \ .
\end{equation}
The coefficients $C_{k,i}$ are given in Table 2 and differ from those in
\cite{ref3} where the older convention \cite{ref17} has been adopted of
counting the gluonic helicity states in $D=4$ rather than in
$D=4+2\ \varepsilon$
dimensions. The latter convention \cite{ref18} is the one chosen to
define all modern NLO parton distributions.

\noindent
Table 2. Coefficients for the expansion of $h_i^g$ in (A6)\\
\begin{tabular*}{\textwidth}{@{~}l@{\extracolsep\fill}llll}
\hline\hline
$i$ & $C_{1,i}$ & $C_{2,i}$ & $C_{3,i}$ & $C_{4,i}$ \\  \hline
$1$ & $4-4(1-\lambda)$ & $\frac{(1-\lambda )z}{1-\lambda z}-1$ & $2$ & $-4$ \\
$2$ & ${8-18(1-\lambda ) \atop +12(1-\lambda )^2}$ & $\frac{1-\lambda}
{1-\lambda z}-1$ & $6\lambda$ & $-12 \lambda$ \\
$3$ & $2(1-\lambda )$ & $0$ & $-2(1-z)$ & $2$ \\ \hline\hline \\
\end{tabular*}

Note that in the limit $\lambda\rightarrow 1\ (m_c\rightarrow 0)$ the
$H_i^{q,g}$ reduce, apart from the obvious collinear logs, to the
massless $\overline{\rm{{MS}}}$ coefficient functions $C_i^{q,g}$
\cite{ref4,ref18}.

\newpage
%
%%%%%%%%%%%%%%%%%%%%%%%%%%%%%%%%%

\newpage
%
%%%%%%%%%%%%%%%%%%%%%%%%%%%%%%%%%%%%%%%%
%
\newpage
\section*{Figure Captions}
\begin{description}
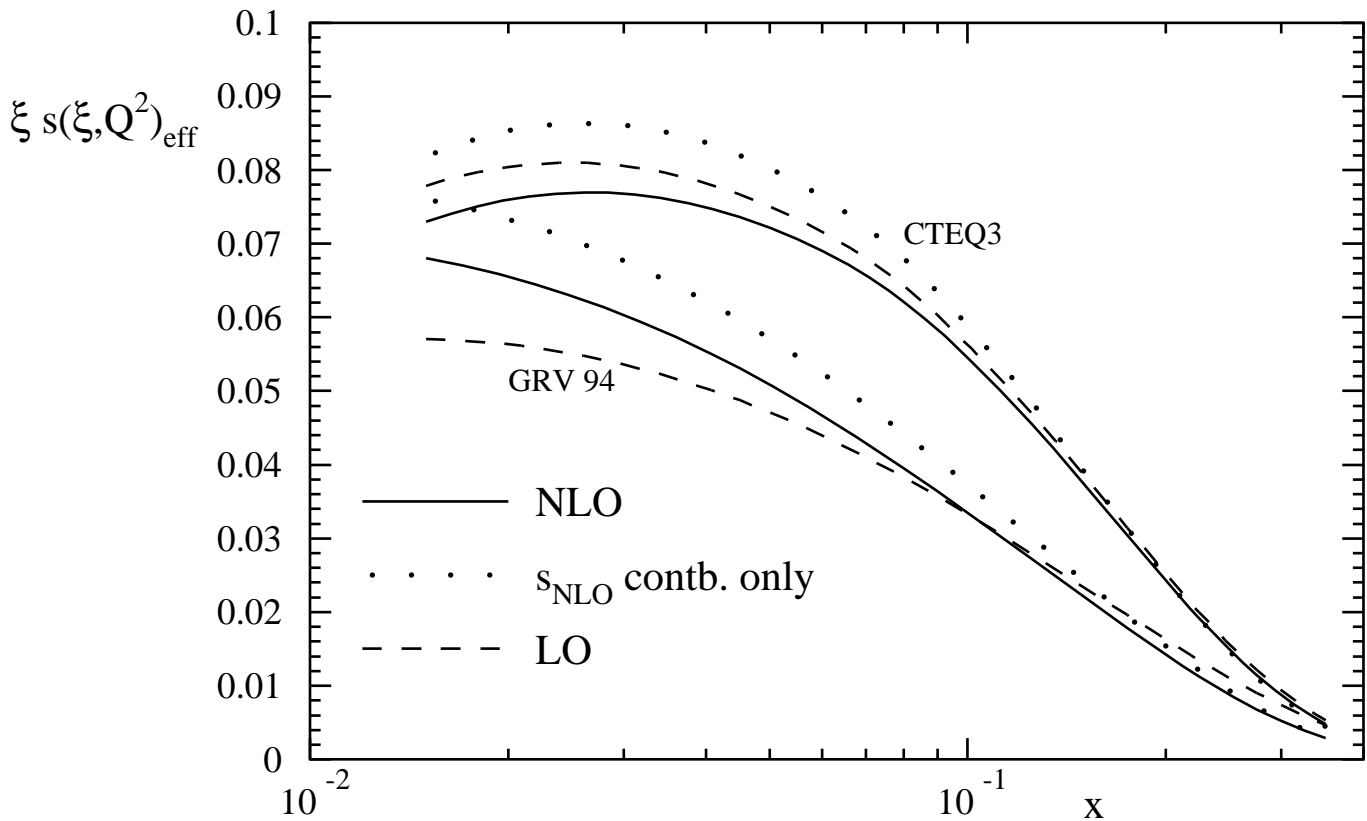
\item[Fig.\ 1] LO and NLO predictions for $\xi s_{eff}$ defined in eq.\
(4), using the GRV \cite{ref4} and CTEQ3 \cite{ref5} parton densities.
The dotted curves refer to using just the NLO strange quark contribution
in eq.\ (2) with all ${\cal{O}}(\alpha_{s})$ terms neglected. Thus the
differences between the dashed and dotted curves illustrate the
differences between the LO and
NLO strange sea densities, respectively. The values of $Q^2$ vary
between $2.4$ to $43.9\ {\rm{GeV}}^2$ according to the experimental
averages \cite{ref8} for $0.015 \le x \le 0.35$ and $E_{\nu}
=192\ {\rm{GeV}}$ \cite{ref7} has been used.
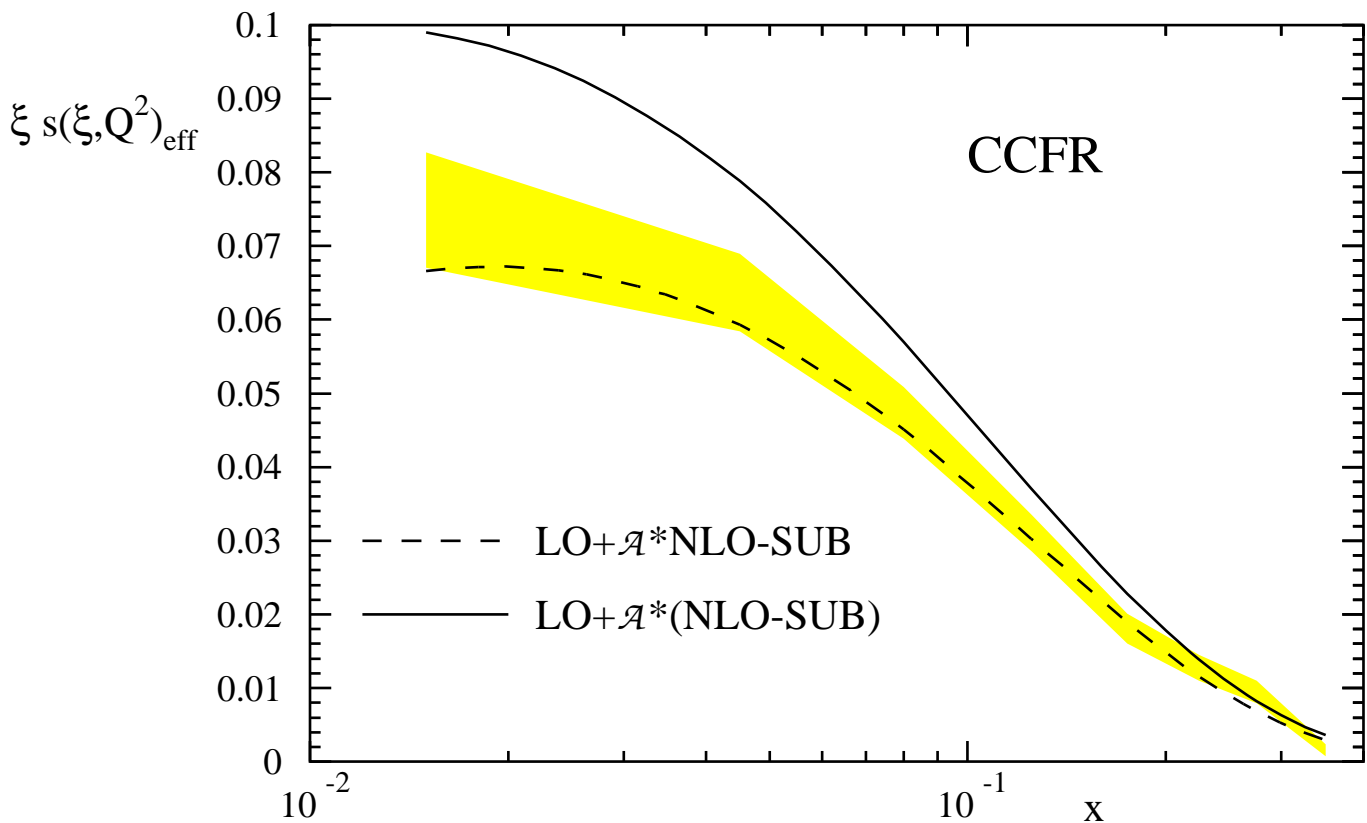
\item[Fig.\ 2] NLO results using the NLO strange sea density of CCFR
\cite{ref7}. The subtraction term (SUB) is defined in (5) and an
acceptance correction factor ${\cal{A}}=0.6$ has been used \cite{ref11}.
The analysis was performed with the original subroutines/matrix elements
of CCFR \cite{ref11}; if the charged current structure functions of GGR
\cite{ref2} are used instead, the results are similar. The shaded area
refers to the CCFR 'data' \cite{ref8}, calculated according to (4'),
where the CKM suppressed contribution in (1) has been subtracted from
the measured full cross section by assuming specific up and down quark
densities \cite{ref7,ref11}. The dashed curve corresponds to the
original CCFR fit analysis \cite{ref7,ref11}.
\item[Fig.\ 3] (a): LO and NLO GRV \cite{ref4} and MRS(A) \cite{ref6}
predictions for $x s(x,Q^2)$ which approximates $\frac{5}{6}F_2^{\nu
N}-3F_2^{\mu N}$ in eq.\ (6). (b): Full NLO result for $\frac{5}{6}F_2^{\nu
N}-3F_2^{\mu N}$ in the FFS (dashed dotted curve) using eq.\ (7), and
the short-dashed curve shows the corresponding result with the $W^+ g
\rightarrow c \bar{s}$ contribution turned off. The factorization scale
chosen is
$\mu^2=Q^2+m_c^2$. The solid curve for $xs$, being the same as in (a),
is shown for comparison. The full NLO MRS(A) result in the 'variable
flavor' scheme is based on eq.\ (8). Both CCFR \cite{ref12} (circles) and
preliminary CCFR \cite{ref13} (squares) $\nu N$ {(Fe--target)} data are
corrected for nuclear shadowing effects, whereas the NMC $\mu N$ data
\cite{ref14} have been obtained from a deuterium target.
\end{description}

\begin{thebibliography}{99}
%
\bibitem{ref1}
M.\ Gl\"{u}ck, E.\ Reya and M.\ Stratmann,
Nucl.\ Phys.\ \underline{B422} (1994) 37.
%
\bibitem{ref2}
M.\ Gl\"{u}ck, R.\ M.\ Godbole and E.\ Reya,
Z.\ Phys.\ \underline{C38} (1988) 441; Erratum \underline{C39} (1988)
590;\\
G.\ Schuler, Nucl.\ Phys.\ \underline{B299} (1988) 21;\\
U.\ Baur and J.\ J.\ van der Bij,
Nucl.\ Phys.\ \underline{B304} (1988) 451.
%
\bibitem{ref3}
T.\ Gottschalk, Phys.\ Rev.\ \underline{D23} (1981) 56.
%
\bibitem{ref4}
M.\ Gl\"{u}ck, E.\ Reya and A.\ Vogt,
Z.\ Phys.\ \underline{C67} (1995) 433.
%
\bibitem{ref5}
H.\ L.\ Lai et al., CTEQ collab.,
Phys.\ Rev.\ \underline{D51} (1995) 4763.
%
\bibitem{ref6}
A.\ D.\ Martin, R.\ G.\ Roberts and W.\ J.\ Stirling,
Phys.\ Rev.\ \underline{D50} (1994) 6734.
%
\bibitem{ref7}
A.\ O.\ Bazarko et al., CCFR collab.,
Z.\ Phys.\ \underline{C65} (1995) 189.
%
\bibitem{ref8}
S.\ A.\ Rabinowitz et al., CCFR collab.,
Phys.\ Rev.\ Lett.\ \underline{70} (1993) 134.
%
\bibitem{ref9}
M.\ A.\ G.\ Aivazis, J.\ C.\ Collins, F.\ I.\ Olness and W.- K.\ Tung,
Phys.\ Rev.\ \underline{D50} (1994) 3102.
%
\bibitem{ref10}
G.\ Kramer, B.\ Lampe and H.\ Spiesberger,
DESY 95-201.
%
\bibitem{ref11}
A.\ O.\ Bazarko, Ph.\ D.\ thesis, Columbia University, Nevis-285 (1994).
%
\bibitem{ref12}
E.\ Oltman et al., CCFR collab.,
Z.\ Phys.\ \underline{C53} (1992) 51.
%
\bibitem{ref13}
CCFR collab. (1993), private communication from R.\ G.\ Roberts.
%
\bibitem{ref14}
M.\ Arneodo et al., NMC,
Phys.\ Lett.\ \underline{B364} (1995) 107.
%
\bibitem{ref15}
B.\ J.\ Edwards and T.\ D.\ Gottschalk,
Nucl.\ Phys.\ \underline{B186} (1981) 309.
%
\bibitem{ref16}
V.\ Barone, M.\ Genovese, N.\ N.\ Nikolaev, E.\ Predazzi and
B.\ G.\ Zakharov,
CERN-TH/95-125 (Z.\ Phys.\ C, to appear).
%
\bibitem{ref17}
G.\ Altarelli, R.\ K.\ Ellis and G.\ Martinelli,
Nucl.\ Phys.\ \underline{B157} (1979) 461.
%
\bibitem{ref18}
W.\ A.\ Bardeen, A.\ J.\ Buras, D.\ W.\ Duke and T.\ Muta,
Phys.\ Rev.\ \underline{D18} (1978) 3998;\\
W.\ Furmanski and R.\ Petronzio,
Z.\ Phys.\ \underline{C11} (1982) 293.
\end{thebibliography}
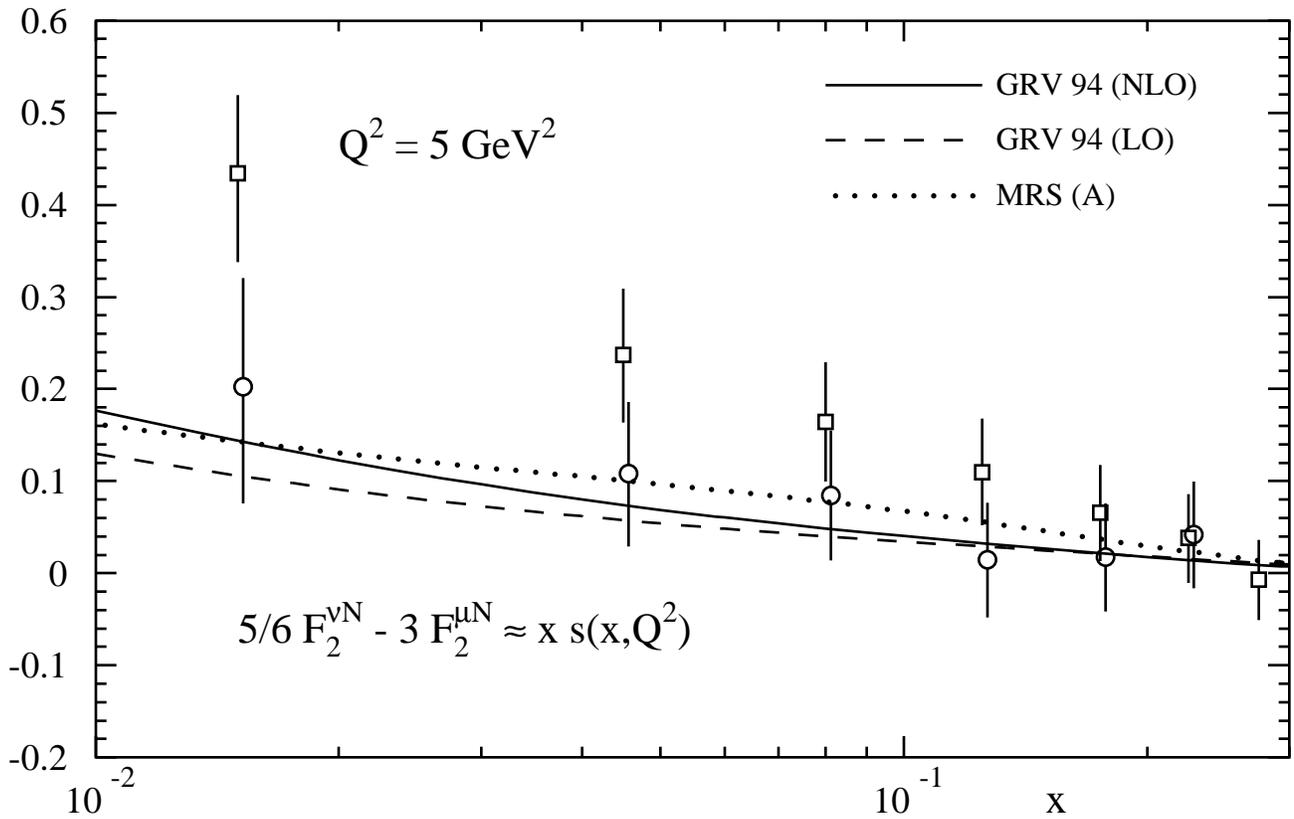
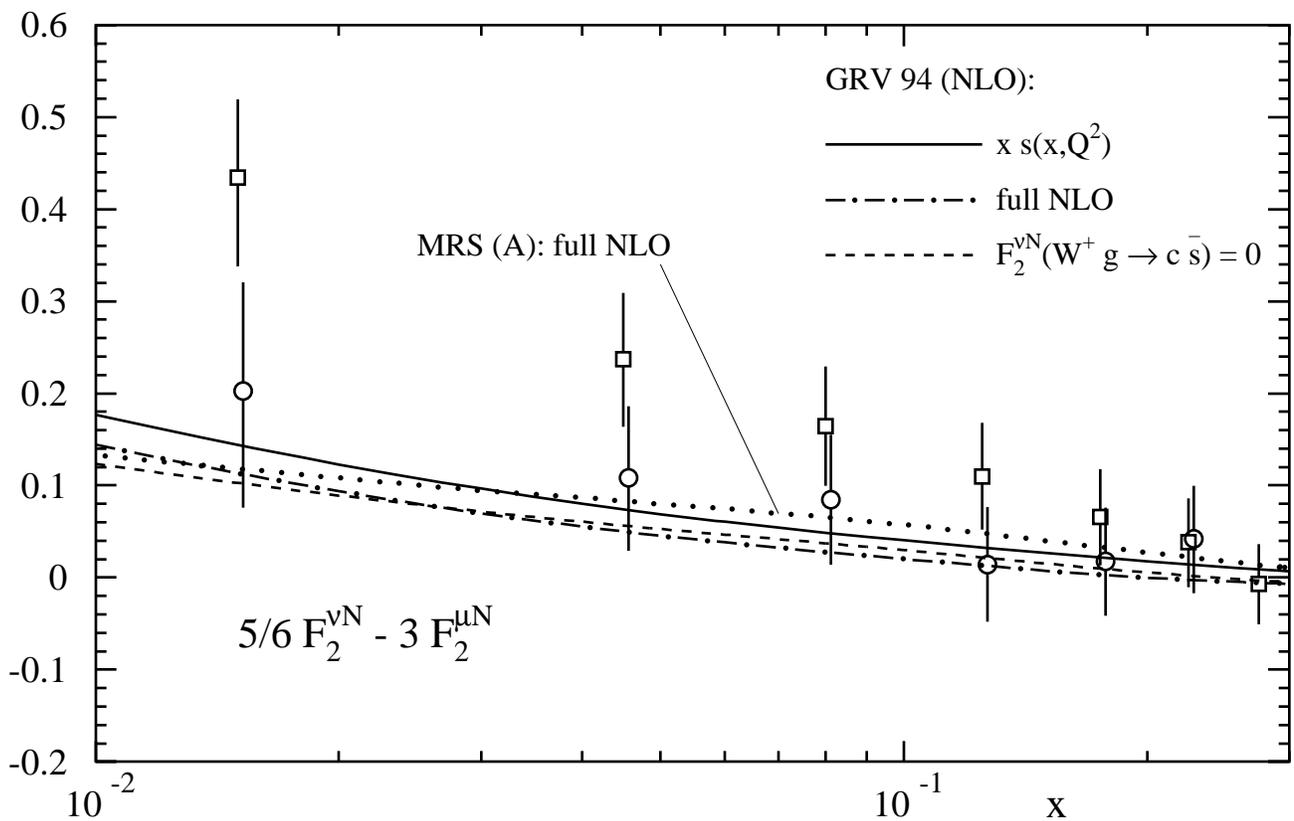
\end{document}